%% file: main.tex
\documentclass[letterpaper,twocolumn,10pt]{article}
\usepackage{usenix2019,epsfig,endnotes}
\usepackage{subcaption}
\usepackage[inline]{enumitem}
\usepackage{url}
\usepackage{array}
\usepackage{amsmath}
\usepackage{tabularx}

\usepackage{etoolbox}
\usepackage{xstring}
\DeclareListParser{\doslashlist}{/}
\newcounter{ndnNameComponentCounter}%
\newcommand{\name}[1]{{%
		\setcounter{ndnNameComponentCounter}{0}%
		\renewcommand{\do}[1]{{%
				\ifnumgreater{\value{ndnNameComponentCounter}}{0}{\allowbreak/}{}%
				\ifnumodd{\value{ndnNameComponentCounter}}{}{}%
				\detokenize{##1}}%
			\stepcounter{ndnNameComponentCounter}}%
		``{\fontfamily{cmtt}\small\selectfont\IfBeginWith{#1}{/}{/}{}\doslashlist{#1}}''%
}}

\usepackage{xcolor}
\newcommand\TODO[1]{}

\begin{document}

%don't want date printed
\date{}

%make title bold and 14 pt font (Latex default is non-bold, 16 pt)
\title{\Large \bf DLedger: An IoT-Friendly Private Distributed Ledger System Based on DAG}

%for single author (just remove % characters)
\author{
{\rm Zhiyi Zhang}\\
UCLA
\and
{\rm Vishrant Vasavada}\\
UCLA
 \and
 {\rm Xinyu Ma}\\
UCLA
\and
{\rm Lixia Zhang}\\
UCLA
} % end author

\maketitle

% Use the following at camera-ready time to suppress page numbers.
% Comment it out when you first submit the paper for review.
\thispagestyle{empty}

\input{0-abstract}

\input{1-intro}
\input{2-related}

\input{2-design}

\input{3-security}

\input{4-sync}
\input{5-sec-assess}

\input{6-imple}

\input{7-discussion}
\input{9-conclusion}
\input{A1-math}

{\normalsize \bibliographystyle{acm}
	\bibliography{main.bib}}

\end{document}

%% file: 0-abstract.tex
\subsection*{Abstract}
\TODO{We should rename PoA to be distinguishable from Proof of Authentication.}

With the ever growing Internet of Things (IoT) market, ledger systems are facing new challenges to efficiently store and secure enormous customer records collected by the IoT devices.
The authenticity, availability, and integrity of these records are critically important for both business providers and customers.
%Over the last few years, cryptocurrencies using blockchain mechanisms, like Bitcoin, have proved that consented and permanent transactions can be kept in a distributed manner.
In this paper, we describe DLedger, a lightweight and resilient distributed ledger system.
%% to enhance security in private business systems.
Instead of a single chain of blocks, DLedger builds the ledger over a directed acyclic graph (DAG), so that its operations can tolerate network partition and intermittent connectivity.
Instead of compute-intensive Proof-of-Work (PoW), DLedger utilizes Proof-of-Authentication (PoA), whose light-weight operations are IoT-friendly, to achieve consensus. 
%for data consensus by using identities.
Furthermore, DLedger is built upon a data-centric network called Named Data Networking (NDN), which facilitates the peer-to-peer data dissemination in heterogeneous IoT networks.

%DLedger takes different approaches from the current practice of using a single blockchain and compute-intensive Proof-of-Work (PoW).
%First, DLedger is based on a directed acyclic graph (DAG), allowing our system to tolerate network conditions like partition or intermittent connection;
%Second, it utilizes Proof-of-Authentication (PoA) instead of compute-intensive PoW, so that the computation recourse is no longer a bottleneck, enabling DLedger to be IoT-friendly.
%Furthermore, DLedger is built upon a data-centric network infrastructure called Named Data Networking (NDN), which facilitate the peer-to-peer data dissemination in heterogeneous IoT networks.

%% file: 1-intro.tex
\section{Introduction}

%In modern business networks, especially in the IoT age, the ledger is of vital importance to keep the customers' data and ensure the steadiness of the business functions.
%According to \cite{evans2011internet}, by 2020, the number of physical objects on the network will reach 50 billion, forming an unprecedented market, and imposing higher requirements on ledger systems to offer better security of data on a massive scale and be more IoT friendly.

In modern business networks, a ledger is of vital importance to keep the customers' data and ensure steadiness of business functions. 
According to \cite{evans2011internet}, by 2020, the number of physical objects in IoT market will reach 50 billion, imposing higher requirements on ledger systems to offer better security of data on a massive scale and be more IoT-friendly.

Today's IoT industry commonly makes use of \emph{centralized} ledger systems.
Our work is inspired by 
%% and previously designed for 
an experimental private solar system~\cite{previous} where the customer energy production/consumption records measured by the rooftop solar gateways were kept on a cloud server for logging.
These records will later be quoted by the business service provider to bill its customers.
However, since a centralized ledger is governed solely by the service provider, this approach is often maligned by customers as well as other financial parties involved in business systems for the following two reasons.
%Today's design for such utility-scale solar system is exposed to the public Internet with a shift from a centralized system to Distributed Energy Resources (DER) which invites cyber attacks.
%For example, the utility control signals and site alarms could be hijacked, damaging the equipments and personnel.
First, the business service provider who controls the server may tamper customer records to obtain unfair advantage.
For example, the solar system service provider may alter the energy usage data to maliciously charge customers who can't prove the existence of original recording data.
Even worse, today, there is little surveillance of such corruption.
%, thus it is increasingly more difficult for system participants to trust the integrity and authenticity of the records and other event logs.
Second, cloud server outages and unstable IoT network connectivity could also probably ruin the data availability.
% when it is needed (e.g., when billing the customers).
% and in IoT scenarios, the unstable network connectivity makes it hard as well for IoT devices to upload the recording data to the remote cloud.

To overcome the drawbacks brought by the centralized solutions, many business providers now pay attention to the blockchain-based \emph{distributed ledger} system.
%For example, the solar network business described should share a distributed ledger system between its rooftop gateway devices so that all the communication regarding customer energy consumption/production, utility signals and site alarms could be logged in consented and permanent way right at the point of its origin providing a decentralized control and surveillance in case of suspect incidents.
However, popular distributed ledgers used in cryptocurrencies, such as Bitcoin~\cite{nakamoto2008bitcoin} and Ethereum~\cite{ethereum}, and other use cases, are not suitable for private IoT business for a number of reasons.
\begin{enumerate*} [label=(\roman*)]
  \item These systems use various resource-intensive gating functions (e.g., proof-of-work, proof-of-stake), which are essentially not acceptable for constrained devices and inefficient for system with massive scale of real-time records.
  
  \item The widely-used blockchain data structure does not suffice with unstable network conditions.
  For example, it is hard for a subnet's blockchain to be integrated into the main chain after a network partition.
  
  \item Moreover, data dissemination and synchronization required by the peer-to-peer (P2P) network often has a low efficiency due to the heterogeneous IoT network. 
  For example, when peers are trying to fetch the missing data in the ledger from the original producer, this producer device could be in sleeping mode and thus not reachable.
\end{enumerate*}

In this paper, we present DLedger, an IoT-friendly distributed ledger system designed for private business networks.
Being data-centric, we build our ledger system over Named Data Networking (NDN)~\cite{ndn}, which is deployable in a private system.
The goal of DLedger is to provide a secure and highly available ledger to keep permanent and immutable data in business contexts.
Throughout DLedger, entities cannot deny or dispute the validity, existence or the ownership of recorded data, thus ensuring the normal business functions.
Addressing the aforementioned challenges, DLedger delivers its design in terms of four main components:
\begin{itemize} [leftmargin=*, itemsep=1pt, parsep=2pt, topsep=4pt]
	\item Records in DLedger are kept in a \emph{Directed Acyclic Graph (DAG)} instead of a single blockchain, allowing simple ledger integration after network partition.
	
	\item DLedger utilizes a compute-efficient gating function called \emph{Proof-of-Authentication (PoA)}, leading to the sufficiency for constrained devices to work.
	Moreover, DLedger provides archiving mechanisms for system participants to reduce the size of their local ledgers.
	
	\item To ensure the robustness of DLedger and prevent potential abuse and attack scenarios, we also propose a set of \emph{security policies}, providing a $(k, N)~k < N$ threshold scheme where unless more than $k$ peers of the total $N$ peers get compromised, the system remains secured.
	
	\item DLedger leverages NDN to achieve effective data dissemination by providing built-in content multicast and resilient data availability.
	Moreover, NDN helps to reduce DLedger's implementation complexity as well.
%	DLedger directly benefits from NDN by leveraging the built-in content multicast and the data-centric security.
%	 which secures data piece itself at the network level without having to adapt costly channel-secure models.
%	Moreover, NDN achieves data replication through inbuilt caching, thus removing the sleeping node problems as data is now location independent and fetched from network as a whole rather than some specific location (IP address).
%    This also eliminates the distinction between light nodes and full nodes as a node doesn't anymore depend on specific full node (IP address).
\end{itemize}

We have implemented a prototype of DLedger and evaluate our system with both theoretical analysis and simulation results.
Our results show that DLedger is of good scalability and robustness, with the ability to mitigate various attack scenarios and potential vulnerabilities.
Throughout the paper, we want to show the power of distributed ledgers to achieving security by combining data openness with verifiable identities.

\paragraph{Organization}
In the rest of the paper, we introduce the related work and key challenges in Section~\ref{sec:motivation} and present the design overview of DLedger in Section~\ref{sec:overview}.
Section~\ref{sec:dag-poa-policy} describes DLedger's ledger design, explaining how DAG, PoW and security policies work together to provide data security.
Section~\ref{sec:network} talks about DLedger's network design.
We assess DLedger system's security in Section~\ref{sec:assessment} and evaluate DLedger in Section~\ref{sec:impl}
We have some discussion about the design in Section~\ref{sec:discussion} and conclude our work in Section~\ref{sec:conclusion}.

%% file: 2-related.tex
\section{Motivation}
\label{sec:motivation}

\subsection{Related Work}

Bitcoin~\cite{nakamoto2008bitcoin} and Ethereum~\cite{ethereum} are two typical and successful examples of distributed ledger systems.
Both are based on blockchain where transactions are kept in blocks and all the blocks are chained together by hash pointers.
To decide who can add the next valid block in the blockchain, miners compete the chance by outpacing each other with higher computation power, leading to immense energy waste because it only allows one game winner and voids all the work done by others.
Another well-known consensus algorithm is Proof-of-Stake (PoS), in which users win the chance of adding the next block by putting their holdings at stake.
%The larger the stake is, the longer duration the stake has, the better chances of winning are.
Many other algorithms are also proposed; for example, \cite{proof-of-space, cryptoeprint:2015:528} propose the Proof-of-Space where peers compete by allocating a non-trivial amount of memory and \cite{proof-of-activity} presents Proof-of-Activity which combines the PoW and PoS.
A cryptocurrency who claims to be IoT-friendly called IOTW~\cite{iotw} proposes a Proof-of-Assignment protocol over the blockchain, where the mining is more lightweight and affordable for IoT devices.
In Proof-of-Assignment, a miner will be selected in two ways.
\begin{enumerate*} [label=(\roman*)]
	\item A centralized server randomly pick up an IoT device.
	\item Each device use their own key and predefined seed to calculate a randomness. A randomness with specific format (e.g., characteristics of n least significant bits) will be selected.
\end{enumerate*}

Blockchain has also gained much attention in other use cases~\cite{conoscenti2016blockchain, zheng2018blockchain}.
As an example, researchers from MIT have developed a decentralized computation platform called Enigma~\cite{zyskind2015enigma}, where a blockchain is maintained between users and service providers to store the access policies to personal data, acting as an access-control moderator.
%Enigma addresses blockchain's privacy issue by placing secret contracts in the blocks and improves the scalability by taking computation from blockchain to powerful off-chain nodes.
To append new blocks, Enigma uses a Proof-of-Stake (PoS) model for worker selection~\cite{enigma-pos}.
Many other works~\cite{vandervort2014challenges, zhang2015iot, Man-IoT-blochain, worner2014your} leverage the unused bytes of the transactions in Bitcoin or Ethereum for their own purposes.
In IoT use cases, for example, \cite{dorri2016blockchain} establishes a distributed trust model employing Beta Reputation System~\cite{josang2002beta} over blockchain in their private smart home devices network, and \cite{yu2018blockchain} proposes an IoT framework based on blockchain and Proof-of-Space.
Besides the distributed ledger systems over TCP/IP, few related works~\cite{blockndn, lou2018blockchain} that are NDN/ICN-based were published as well.
For example, \cite{blockndn} presents a Bitcoin-like blockchain system over NDN called BlockNDN.
% and another work~\cite{lou2018blockchain} proposes the use of blockchain to manage the digital keys in the NDN.
%, but does not provide details answering how consensus is achieved in the blockchain.

%
%As observed, a lot of existing solutions are IoT friendly since most of them are developed over a crude blockchain technology without tweaking it to fit IoT needs, such as network partition friendly, lightweight gating control mechanism, and smart storage/archiving.

Specifically, we identify the works of IOTA~\cite{iota}, Nano~\cite{nano}, and Byteball~\cite{byteball}, which are graph-based distributed ledger systems.
IOTA~\cite{iota} is a cryptocurrency system which claims to be IoT friendly given its feeless transaction and lightweight PoW.
IOTA's distributed ledger is built upon an underlying data structure called tangle~\cite{tangle}, derived from Directed Acyclic Graph (DAG).
Each node in the tangle is a block carrying the transactions made with IOTA coins, and similar to chained blocks in blockchain, blocks are also associated by hash references.
IOTA allows each system participant to append new blocks by verifying two existing blocks and doing a lightweight PoW.
To select blocks to approve, IOTA utilizes a weighted random selecting algorithm called Markov Chain Monte Carlo (MCMC) to walk from ancient blocks towards the latest ones.
Once a record has been referred by enough blocks, it is considered that the system has achieved the consensus on this record.
IOTA utilizes tangle and thus allows multiple transaction to be processed in parallel, vastly improving the system capacity compared with blockchain-based systems.
Nano~\cite{nano} and Byteball~\cite{byteball} are also cryptocurrencies utilizing graph-based ledgers.
Nano, similar to IOTA, uses PoW as anti-spam protection while Byteball gets rid of PoW and achieves consensus by forming a single chain called ``main chain" within the graph.
This main chain consists of the blocks selected by trusted third-party users.

\subsection{Observations and Key Challenges}
\label{sec:observation}

As observed, most consensus protocols are relying on system participants' resources (power of calculation or storing), which is incompatible with the constrained resources of constraint IoT devices due to ``no muscle to show".
Consequently, instead of directly participating in the ledger system, IoT devices need to wait for powerful miners to add their transactions into the blockchain.
Among the protocols, Proof-of-Assignment~\cite{iotw} seems to be IoT-friendly.
However, the first minor selection method relies on the central server, potentially leading to service provider's malicious intervention and single-point-of-failure, while the second selection method can easily be compromised because attackers can simply hire computation power (not too much because IOTW's mining is so lightweight) and work out a randomness to win the selection.

Moreover, the single blockchain structure is not IoT-friendly as well for the following reasons.
First, it limits the system capacity because there is always only one next record, forbidding the parallel efforts.
Bitcoin is such an example where each block takes about 10 minutes to be appended into the chain and the maximum throughput is 3.3--7 transactions/sec~\cite{croman2016scaling}.
Also, in case of network partition, each subnet will keep appending new blocks to their view of chain, thus leading to two different versions of blockchains that are not compatible.

Compared with blockchain-based systems, IOTA utilizes tangle and thus allows multiple transaction to be processed in parallel, vastly improving the system capacity.
However, IOTA is not suitable as a distributed ledger in IoT network as well.
On the one hand, IOTA's lightweight PoW is still heavy for constrained IoT devices because it can take minutes to compute one PoW, potentially influencing the normal IoT functions.
On the other hand, since a miner is anonymous to the system, it can secretly lease out PoW computation to the modern computers for whom PoW is not so resource-consuming, thus able to append many blocks to the system in a short time.
This allows them to not only make self-approvals where a peer can keep referring one's own (even invalid) blocks but also make system accept them by controlling all the tips.
Moreover, since the hash pointers are made from newer blocks to old blocks while MCMC requires the random walk from old to new, it leads to inefficiency because finding successive blocks is non-trivial. In the worst case, a peer even needs to iterate the whole graph for each step.

Other graph-based distributed ledgers are also not suitable as a private ledger used in IoT.
For example, in Nano, the PoW could be computed in order of seconds by modern computers, thus the system may subject to transaction flooding when attackers are with computation power.
Regarding Byteball, though based on graph, the consensus of data is still in a singe chain, which is insufficient in network partition.
Byteball argues that the network partition is almost impossible in the public Internet, but in IoT network, this assumption does not hold anymore.

Therefore, designing a distributed ledger system for IoT business model requires changing in both ledger design and network design.
In particular, we identify three main challenges to address in our work.

\textbf{1. Providing robust ledger whose data consensus is resilient to unstable IoT network conditions.}
In the case of network partition or intermittent connectivity, entities from different subnets should still be able to contribute to the ledger system.
After network failure, the distributed ledger can quickly recover from the partition by aggregating the data generated by different subnets.

\textbf{2. Working with constrained capacity of IoT devices and the massive scale of data.}
The distributed ledger should be efficient enough for constrained devices to append their own data and at the same time, preventing the potential abuse and attack scenarios.

\textbf{3. Filling the gap between inefficient data dissemination in IoT and the high data throughput required by the P2P network.}
The ledger should support efficient data dissemination for the routine synchronization among peers.

%% file: 2-design.tex
\section{DLedger Design Overview}
\label{sec:overview}

\textbf{Goals and Design Decisions}
DLedger aims to keep non-repudiable and permanent records in a distributed ledger system that can work with resource-constrained IoT devices.
To achieve the goal and address the challenges identified, DLedger takes four design decisions:
\begin{itemize} [leftmargin=*, itemsep=1pt, parsep=2pt, topsep=4pt]
    \item DLedger utilizes DAG to keep the recording data.
    \item DLedger uses PoA as the gating function to allow entities adding new records.
    \item DLedger ensures system robustness by a set of security policies.
    \item DLedger is built over a data-centric network.
\end{itemize}

\noindent\textbf{Assumptions}
In DLedger, each entity is a valid customer node, or a device deployed by the business provider or partners involved who are authorized to access the records.
Our distributed ledger system is based on the underlying assumption that entities in the private system have established the \emph{trust relationship} with the help of the \emph{identity manager} deployed by the business provider.
To be specific:
\begin{enumerate*} [label = (\roman*)]
    \item Each entity trusts the \emph{identity manager}, i.e., install the identity manager's digital certificate and is able to verify the signature directly or indirectly signed by the identity manager.
    \item Each entity in the system has an identity name and each has been issued a digital certificate by the designated identity manager in advance to bind the entity's identity name with the public/private key pair.
\end{enumerate*}
There can also be more than one identity managers in multi-party business model; in that case, cross-certifications among the identity managers are required as well.

\begin{figure} [t]
	\centering
	\includegraphics[width= 0.85\linewidth]{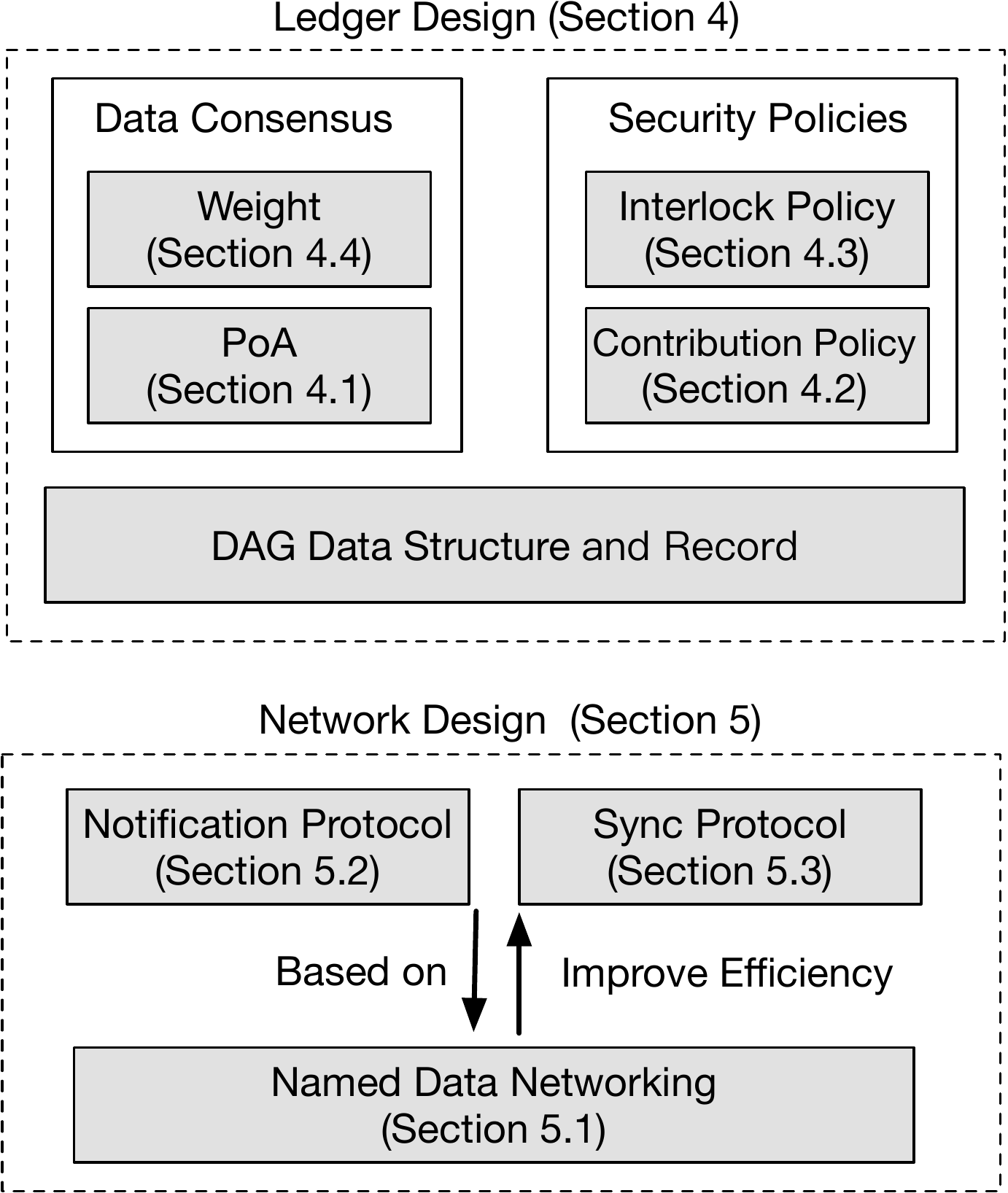}
	\caption{DLedger's design structure}
	\label{fig:components}
\end{figure}

An overview of the DLedger's structure is shown in Figure~\ref{fig:components}.
In a nutshell, DLedger is designed to facilitate security solution development by recording and interlocking every record in the system, creating and maintaining an immutable and distributed ledger.
All entities in a DLedger system, including the identity manager, together form a peer-to-peer (P2P) network. 
Each peer appends its new records into the DLedger after making \textit{approvals} to other peers' records after verifying their validity.
The identity manager also appends entity certificates issuance and certificate revocation records into the ledger.
At the moment, the distributed ledger stores two types of record content:
\begin{itemize} [leftmargin=*, itemsep=1pt, parsep=2pt, topsep=4pt]
	\item entity's application records, e.g., solar energy usage
	\item identity management operations which include certificate issuance and revocation.
\end{itemize}
The consensus on a record in the distributed ledger is achieved if enough number of entities has approved this record.
The number of approvers is called the \emph{weight} of a record in DLedger.

\begin{figure} [ht]
	\centering
	\begin{subfigure}{.25\textwidth}
		\centering
		\includegraphics[width= \textwidth]{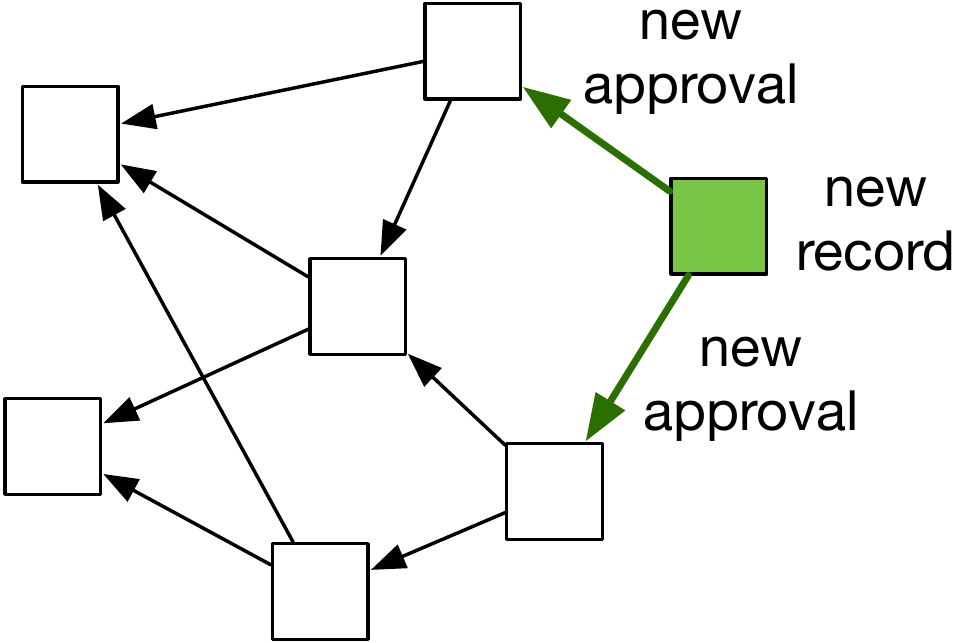}
		\caption{A DAG}
		\label{fig:dag}
	\end{subfigure}
	~
	\begin{subfigure}{.18\textwidth}
		\centering
		\includegraphics[width=\textwidth]{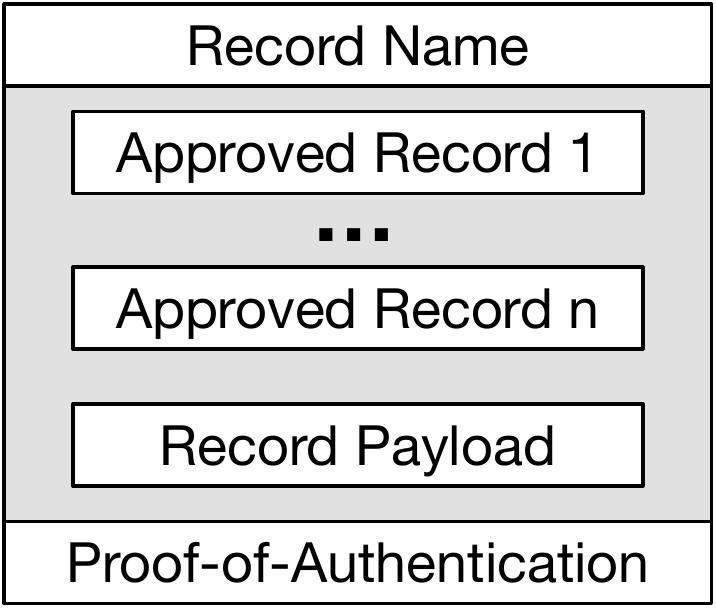}
		\caption{Record Format}
		\label{fig:record}
	\end{subfigure}
	\caption{DLedger DAG and Record}
\end{figure}

Sharing the similar idea as the IOTA's tangle, DLedger stores all records in a Directed Acyclic Graph (DAG) instead of a single blockchain.
As shown in Figure~\ref{fig:dag}, each vertex represents a record carrying its payload data while each edge represents an \textit{approval} made by a record to another.
In DLedger's DAG, a newly generated record (vertex) will be placed adjacent (chained) to \textit{n} ($n \geq 2$) existing records in the DAG, establishing the approvals from the new record to the previous ones.
Figure~\ref{fig:dag} shows an example of DAG when $n=2$.
\textit{Tailing} records are those records not approved by any other records in the DAG.

When a peer $P$ generates a new record $R$, the peer takes the steps as follows.
\begin{enumerate} [leftmargin=*, itemsep=1pt, parsep=2pt, topsep=4pt]
    \item $P$ randomly selects $n$ tailing records generated by other peers. 
    
    \item It will then verify the validity of these records and the records directly or indirectly approved by these records.
    If any of the $n$ records is invalid, i.e. the record is malicious or it approves an invalid record, $P$ should drop it and repick another.
    
    \item The peer then adds the names of the $n$ selected records and the application payload into the new record $R$ as shown in Figure~\ref{fig:record} and sets $R$'s name to:
    \begin{center}
        \name{/DLedger/<Generator ID>/<Record Hash>}
    \end{center}
    where \name{<Generator ID>} is $P$'s unique ID and \name{<Record Hash>} is the digest calculated over $R$.
	
	\item Finally, $P$ appends a \emph{Proof-of-Authentication (PoA)} by digitally signing the record, making its authenticity and integrity verifiable.
\end{enumerate}
%The signature of a record that can be verified with the public key certified by the identity manager is called \textit{Proof-of-Authentication (PoA)}, used as a gating control mechanism in DLedger as an analogy of the Proof-of-Work in cryptocurrency distributed ledgers such as Bitcoin.
%If a record satisfy the security polices and has been verified by enough number of peers, the record is considered to be confirmed by the whole system and we say the P2P network has achieved the consensus on this piece of data.

The peer then utilizes the new record notification protocol (Section~\ref{sec:network:advertisement}) to advertise it to the whole P2P network.
After receiving the notification, other entities will fetch the record and add it to their local ledger after authentication.
There also exists a ledger synchronization process (Section~\ref{sec:network:sync}) triggered both periodically and after out-of-sync events (e.g., sleeping mode, network failures, etc.).
Leveraging the notification protocol and the synchronization protocol, DAG is replicated and synchronized among all entities.

A fundamental difference that distinguishes our work from existing solutions is the data-centric implementation over NDN.
Dissimilar to TCP/IP network, NDN make the named and secured data as the thin waist of the network.
In NDN, an application fetches data by sending an \emph{Interest} packet carrying the desired data name.
The network forwards the Interest by its name and fetches the corresponding \emph{Data} packet from its origin or an in-network cache.
We elaborate the details of these two protocols and their data-centric features in Section~\ref{sec:network}.

\begin{figure} [ht]
	\centering
	\includegraphics[width= 0.8\linewidth]{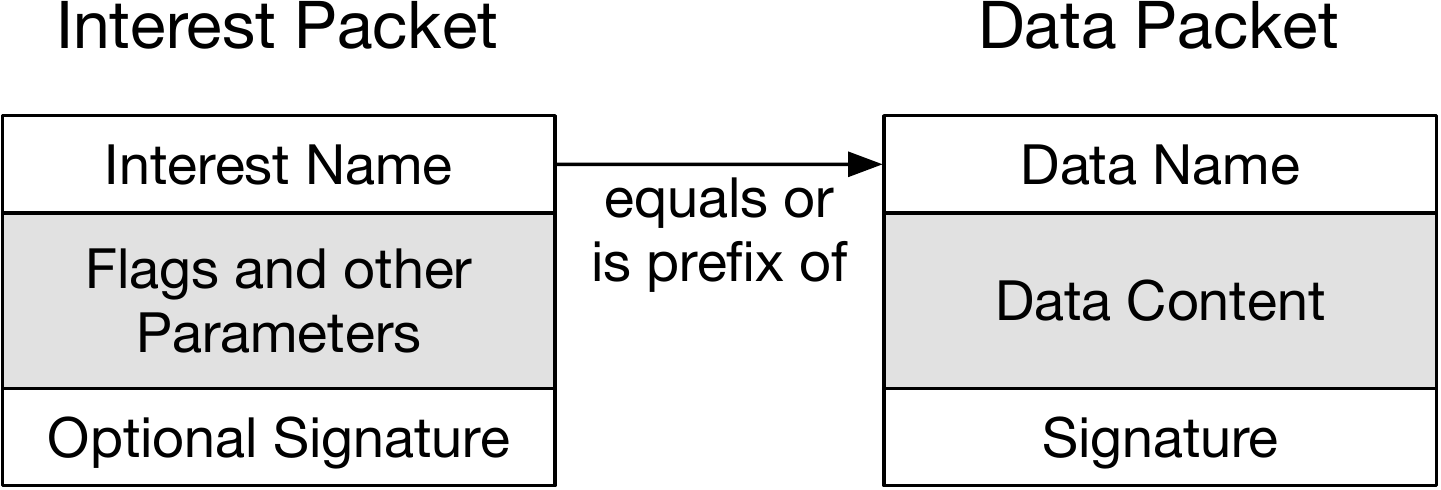}
	\caption{Interest and Data packet in NDN}
	\label{fig:ndn-packet}
\end{figure}

DLedger adopts a set of security policies for record receivers to check the incoming records, preventing potential threats such as spam attack and collusion attack, and misbehaving operations like laziness (contributing nothing to system security).
These security policies assures a $(k, N)~k < N$ threshold scheme where unless more than $k$ peers of the total $N$ peers get compromised, the system remains secured.

\paragraph{Incentives for Entities}

The incentive for a peer $P$ to contribute is the dependency of a record's validity on its approvals. Thus, actively approving records leads to the confirmation of the records $P$ generates and benefits its data authenticity, integrity, and availability.
By contrast, a passive peer may attach its records to invalid records and thus make them rejected by the loyal majority of entities, losing guarantee of the data security.

%It is possible that ``lazy" entities may want to make approvals to the already confirmed records to escape the validation work.
%To prevent such laziness, as shown in Section~\ref{sec:security:policies}, DLedger leverages the contribution policy and force entities to contribute to the system.

The motivation of keeping one's local ledger up-to-sync is to get the latest status, which is important to protect its own interest.
For example, the identity manager may insert the certificate revocation records into the distributed ledger.
Without noticing such records, an entity may wrongly approve a record signed by a revoked key, resulting in generation of abandoned records. 
Moreover, keeping the freshness of the local ledger helps entities to know whether their own records get accepted, confirmed, or abandoned by the whole system.
%\TODO{Vishrant: another motivation: syncing lets them debug why and if their record was abandoned?}

%% file: 3-security.tex
\section{DAG-Based DLedger with Interlocked PoA-Verifiable Records}
\label{sec:dag-poa-policy}

This section presents the detailed design of the ledger system in terms of 
\begin{enumerate*} [label=(\roman*)]
	\item how PoA works,
	\item how DLedger ensures the success of recording functionality by security policies (i.e., contribution policy and interlock policy),
	and
	\item how the system achieves consensus on the recorded data. 
\end{enumerate*}
We also describe a mechanism for storage-constrained IoT device to reduce required storage of the local ledger.

\subsection{Proof-of-Authentication}
\label{sec:poa}

In DLedger, instead of computing crypto puzzles or allocating other resources, entities append a PoA to the record, associating it to a specific entity, so its creator cannot refuse to admit the existence and ownership of it.
To verify a PoA, an entity confirms:
\begin{itemize} [leftmargin=*, itemsep=1pt, parsep=2pt, topsep=4pt]
	\item The public key, which the identity manager issued to the creator, can verify the PoA.
	\item The record containing the key's certificate must be confirmed.
	\item No confirmed record has revoked the certificate.
\end{itemize}
PoA is a difference enabler in our ledger design because it allows the system to distinguish records generated by different entities.
As described later in this section, the interlock policy and the consensus process are all based on this attribute.

PoA provides the authenticity and integrity of the data but does not suffice to prevent evil entities from abusing the ledger.
One reason for PoA is its lightweight. To generate numerous PoA in a short time is not difficult even for constrained devices, given the wide usage of ECC chips like ATECC508A in IoT.
Therefore, DLedger adopts security policies to force entities behave.

\subsection{Contribution Policy}
\label{sec:policy:contribution}

The contribution policy enforces peers to verify unconfirmed records, to be more specific, tailing records when creating a record.
This policy enhances robustness in the following aspects.
\begin{itemize} [leftmargin=*, itemsep=1pt, parsep=2pt, topsep=4pt]
	\item Ensure the continuous verification and confirmation of new records, preventing the escalation in the size of unconfirmed records.
	\item Preventing lazy peers who approve old records and contribute nothing to security.
\end{itemize}
Whenever a \emph{tailing record} arrives, it is checked against the contribution policy, e.g., after fetching a new record advertised by another entity.

When a new record $R$ approving $n$ existing records is received, ideally these $n$ records should be tailing records:
\begin{equation} 
\label{eq:old-contribute}
\forall i \in range(1, n) ~~ w_i = 0 
\end{equation}
where $w_i$ is the weight of $i$th approved record.
However, network delay and concurrent approvals make it possible for them to be no longer tailing records upon arrival.
To avoid unexpected rejections, we introduce a threshold $W_{contribution}$ w.r.t.:
\begin{center}
	$0 < W_{contribution} < W_{confirm}$
\end{center}
When verifying a new record, instead of using rule~\ref{eq:old-contribute}, the peer checks whether it satisfies:
\begin{equation}
\label{eq:contribute}
\forall i \in range(1, n) ~~ w_i < W_{contribution}
\end{equation}
The safe choice for an entity is to randomly select $n$ tailing records to approve whenever generating a record, but in the case of insufficient tailing records, recent records with a weight not exceeding 
$W_{contribution}$ are also allowed to select.

For any record violating the condition, unless more than $W_{confirm}$ entities accept it ignoring the policy, it will never gain enough weight to be confirmed. Therefore, it forms the $(k, N), ~~k = W_{confirm}$ scheme. Similar statements hold for the other policy.

\subsection{Interlock Policy}
\label{sec:policy:interlock}

The interlock policy prevents a peer $P$ approving records generated by itself.
In other words, no adjacent two records should be generated by the same peer:
\begin{center}
	$\forall{R_i, R_j} ~~ R_i$ and $R_j$ are adjacent $\Rightarrow P_i \neq P_j $
\end{center}
Any arriving records against the condition above will be rejected, no matter whether it is a tailing record or not.

Similar to the contribution policy, we have the scheme $(k, N), ~~k = W_{confirm}$ hold.

\begin{figure} [ht]
	\centering
	\includegraphics[width= 0.95\linewidth]{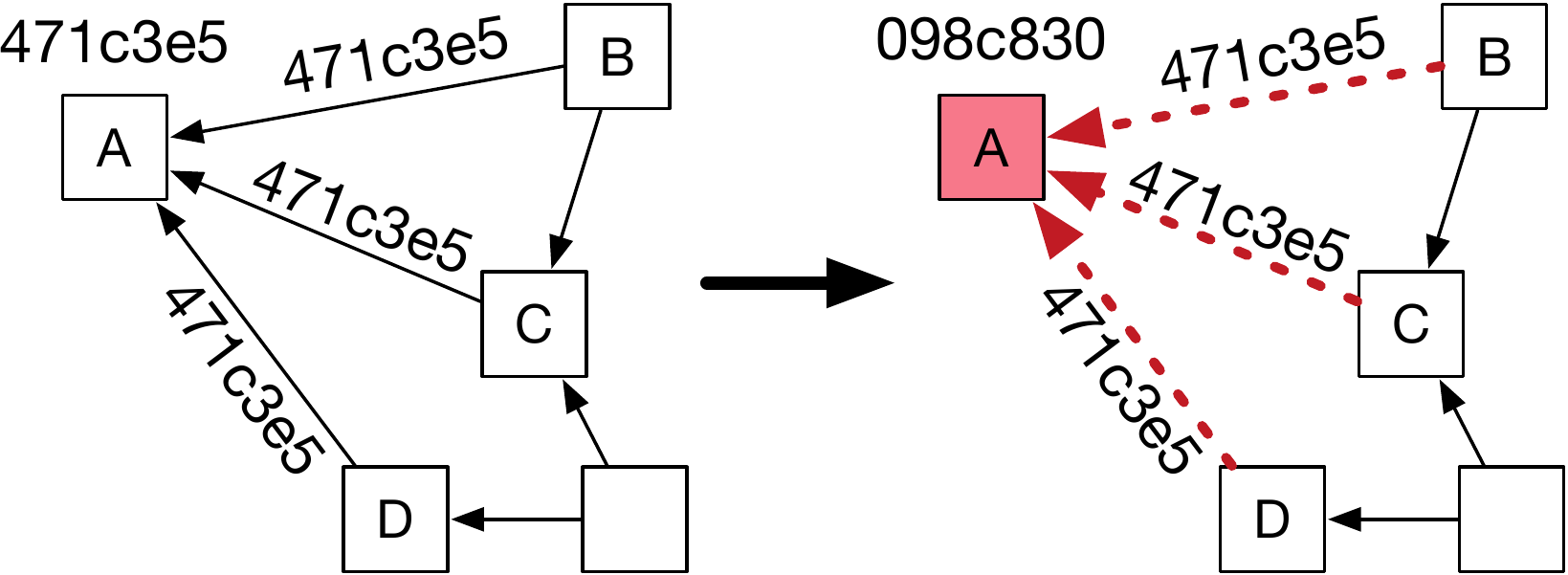}
	\caption{Modifying Records in DLedger}
	\label{fig:modify}
\end{figure}

The interlock policy forces records from different entities to interlock each other by approval---refering to names.
As shown in Figure~\ref{fig:modify}, since each record name contains the hash value of the block, modifying a record (record $A$ in the figure) will void all the approvals made to it (avalanche effect in fact).
Note that these approvals are protected by PoAs, and per the interlock policy, $B$, $C$, and $D$ were generated by different entities from $A$'s.
Therefore, no one can modify a record without knowing all other entities' private keys.

Moreover, the interlock policy limits that any peer can only increase the unconfirmed records' depth by at most one, hence eliminating the attack scenarios where a compromised entity endlessly append new records, add to the unconfirmed depth and burden other peers.

%\subsection{Definition of Record Validity}
\subsection{Consensus Through Record Weight}
\label{sec:design:append}

\begin{figure} [ht]
	\centering
	\includegraphics[width= 0.7\linewidth]{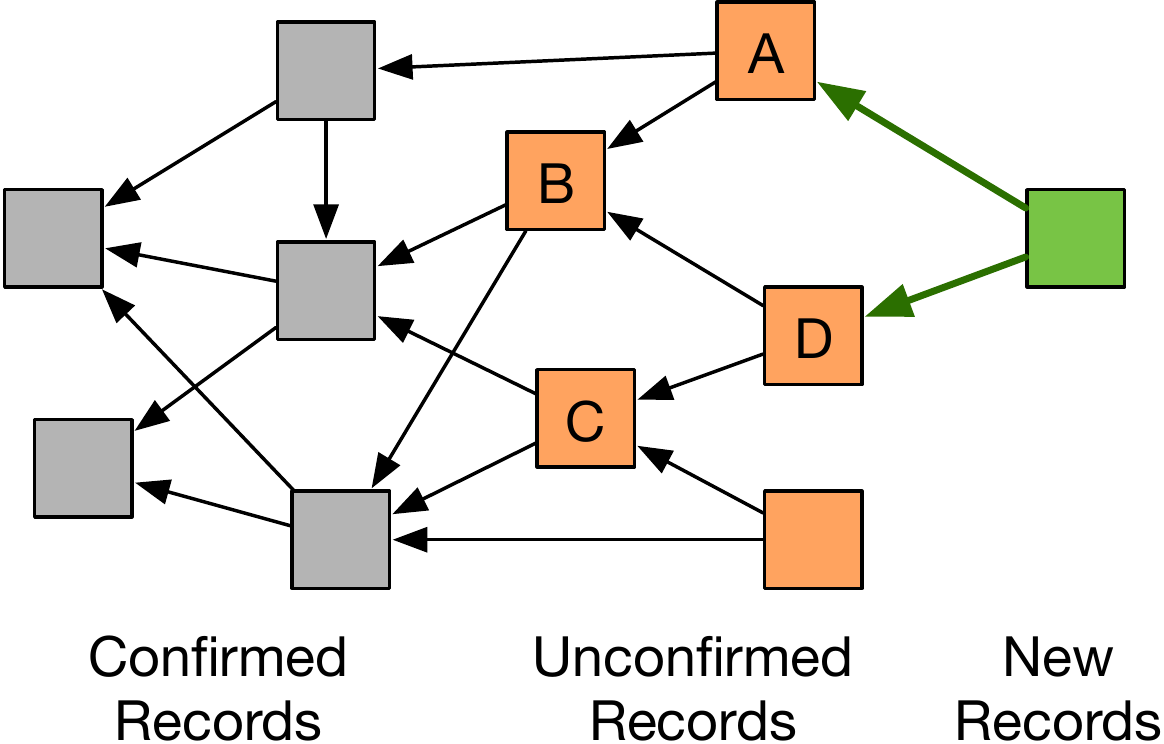}
	\vspace{1mm}
	{\footnotesize \\ To approve record $A$ and $D$, the entity needs to validate record $A$, $B$, $C$, and $D$. \par}
	\caption{Approving the validity of existing records}
	\label{fig:approve}
\end{figure}

When an entity approves a record, it acknowledges the its validity and the records it approved directly or indirectly until reaching confirmed records.
Figure~\ref{fig:approve} shows an example of such process in case when $n=2$.
To be more specific, this entity has to verify that all records involved satisfy the following four requirements.
\begin{enumerate} [leftmargin=*, itemsep=1pt, parsep=2pt, topsep=4pt]
    \item They carry valid PoAs.
    \item The tailing records cannot approve a record whose weight is too large by the \emph{contribution policy}.
    \item A record cannot approve another record generated by the same entity given the \emph{interlock policy}.
    \item The payload carried in these records satisfy the application-level semantics.
\end{enumerate}
The first three requirements are defined by DLedger while the last one is controlled by the application layer.
A typical example of the application-level rule is that to verify a Bitcoin transaction, peers need to trace back the transaction history of the buyer to ensure it has enough coins for the purchase.
In this paper, we focus on the distributed ledger and network solutions, leaving the application-level rules to the specific use cases.

\begin{figure} [ht]
	\centering
	\includegraphics[width= 0.8\linewidth]{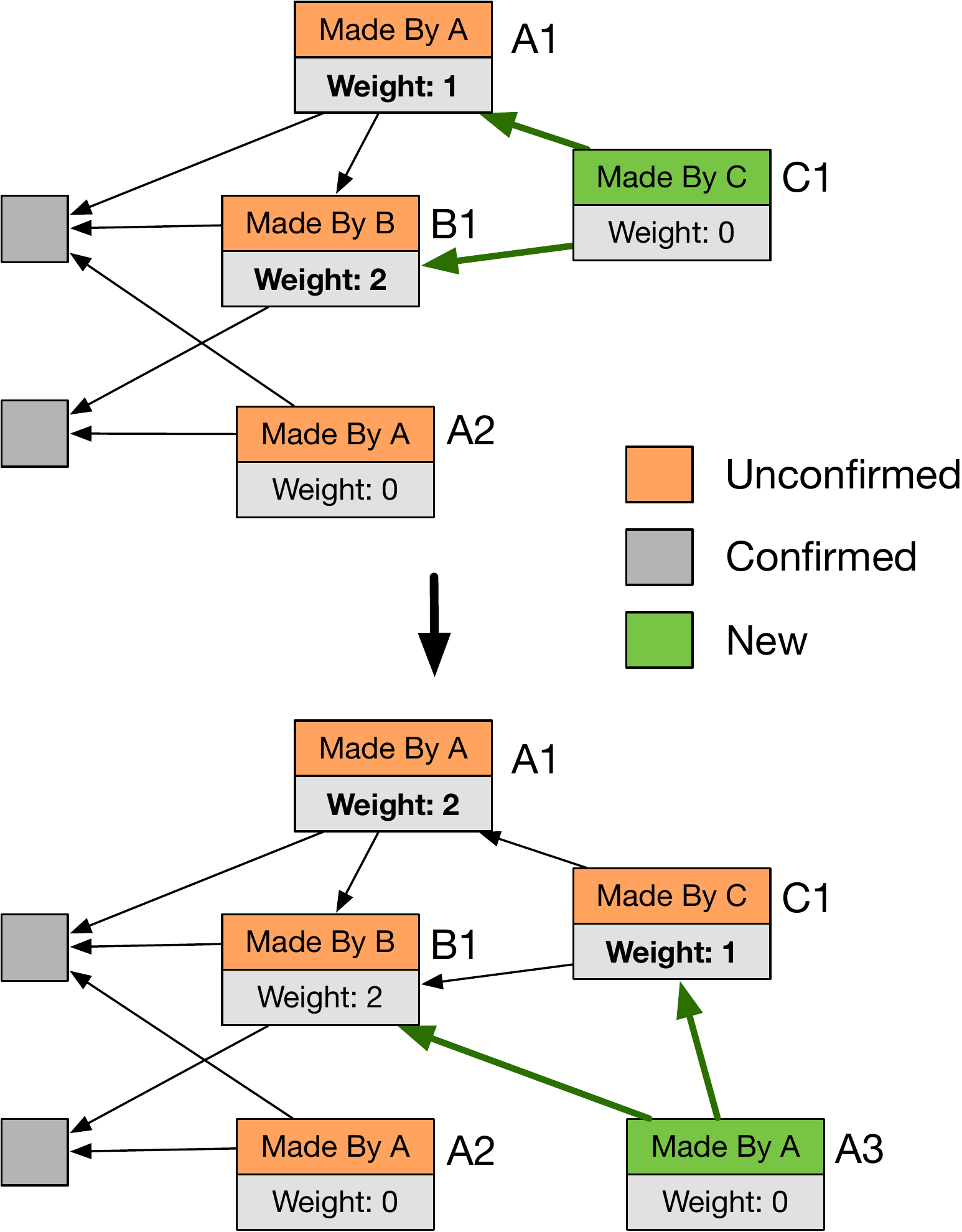}
	\medskip
	
	{\footnotesize The appending of $C1$ increase the weight of $A1$ and $B1$ by 1. Then, the insertion of $A3$ will add to the weight of $C1$ and $A1$ due to the direct and indirect approval respectively, but not $B1$ because it has been already approved by $A$.\par}
	\caption{An example of weight increase}
	\label{fig:weight}
\end{figure}

The DLedger system achieves consensus through record \emph{weight}.
The weight of an record is the number of entities approving it directly or indirectly, denoted by $w$ here.
A newly generated record have a zero weight. 
After being advertised and spreading across the P2P network, it gets approved accompanied by a cumulative increasing in its weight.
Figure~\ref{fig:weight} shows a simple example.
Records are categorized by whether their weight exceed a threshold $W_{confirm}$ into two states:
\begin{itemize} [leftmargin=*, itemsep=1pt, parsep=2pt, topsep=4pt]
	\item \emph{Unconfirmed record}: $w < W_{confirm}$.
	\item \emph{Confirmed record}: $s \geq W_{confirm}$.
\end{itemize}

An eventual goal for a record is to gain enough weight to become confirmed. 
If it cannot, however, it may be voided after an application-defined period.
By contrast, a confirmed record means the system has achieved consensus on the validity of it and will permanently store it. 

\subsection{Reducing Storage of Local Ledger}
\label{sec:design:archive}

The monotonical appending of records continues consuming the space. 
To fit DLedger into storage-contrained IoT devices, when the provisioned space approaches exhaustion, an entity can reduce the size of the local DAG by
\begin{enumerate*} [label=(\roman*)]
	\item pruning all the records that stay unconfirmed after a period of time,
	and
	\item taking out the ancient confirmed records that are far away from the tailing records.
\end{enumerate*}
Regarding the ancient confirmed records taken out from the DAG, an entity can has following options depending on specific use cases.
\begin{itemize} [leftmargin=*, itemsep=1pt, parsep=2pt, topsep=4pt]
	\item The entity can transfer it to some storage service (e.g., cloud server) for backup.
	\item These records can be dropped if there are designated nodes deployed by service provider to store the full copy of the DAG.
	Since the records are interlocked and protected by PoA, the service provider cannot modify the content in the existing DAG.
	\item Without backup service available and designated nodes deployed, it is still safe for the entity to drop the records if these records are no longer useful.
	Taking the solar gateway system as an example, after the service provider bills the customer rightly, old records in each entity's local ledger become droppable.
\end{itemize}

Note that this process takes place at each individual entity with no multi-party coordination.

%% file: 4-sync.tex
\section{DLedger over a Data-centric Network}
\label{sec:network}

Two basic network functions are required in DLedger: \emph{notification} and \emph{synchronization}, posing a big challenge to heterogeneous IoT networks which lack continuous connectivity.
Notification enables entities to advertise new records, while synchronization makes a node's local state consistent with others.
An entity is not guaranteed to be always online; a record is not guaranteed to be always accepted.
Consequently, a peer needs to synchronize its DAG with the P2P network both on a timely basis and after a known failure, such as link failure, device failure, low-power mode, etc.

This section introduces the concepts of NDN and describe how DLedger exploits NDN protocols to facilitate data dissemination in an IoT network and simplifies the implementation.
In particular, DLedger leverages NDN's data-centric features in three aspects.
\begin{itemize} [leftmargin=*, itemsep=1pt, parsep=2pt, topsep=4pt]
	\item 
	DLedger utilizes in-network caching and built-in content multicast to  improve the data transmission efficiency especially when the network  is unstable.
%	 since every NDN node is cache-capable at the network layer. 
	\item 
	The wireless multicast in IoT network can benefit from the more accurate packet suppression mechanism because NDN provides application semantics at the network layer.
	\item
	DLedger's implementation can be markedly simplified by developing it over a data-centric network.
\end{itemize}

\begin{figure} [ht]
	\centering
	\begin{subfigure}{.2\textwidth}
		\centering
		\includegraphics[width= \textwidth]{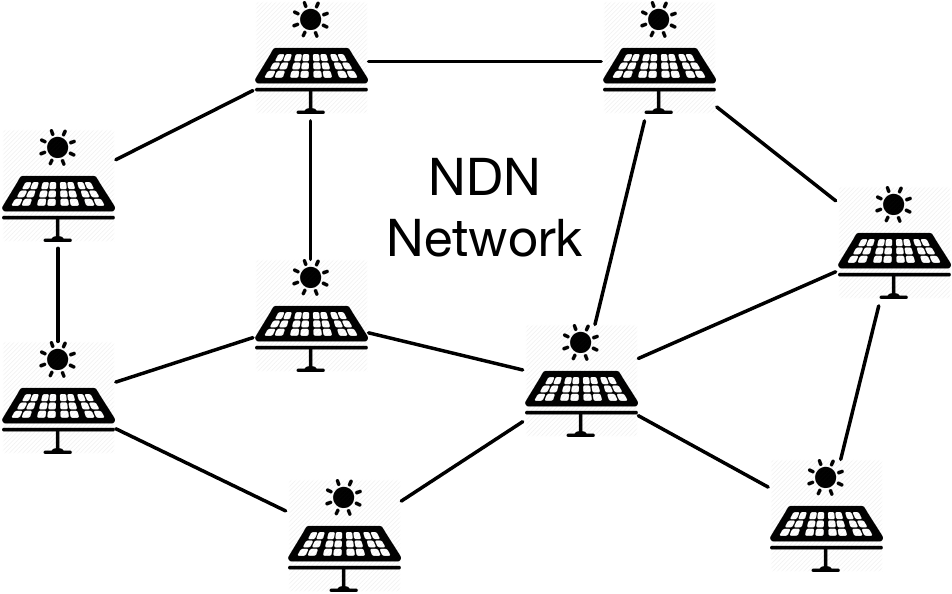}
		\caption{Wireless mesh NDN network}
		\label{fig:mesh}
	\end{subfigure} \hspace{4mm}
	~
	\begin{subfigure}{.2\textwidth}
		\centering
		\includegraphics[width=\textwidth]{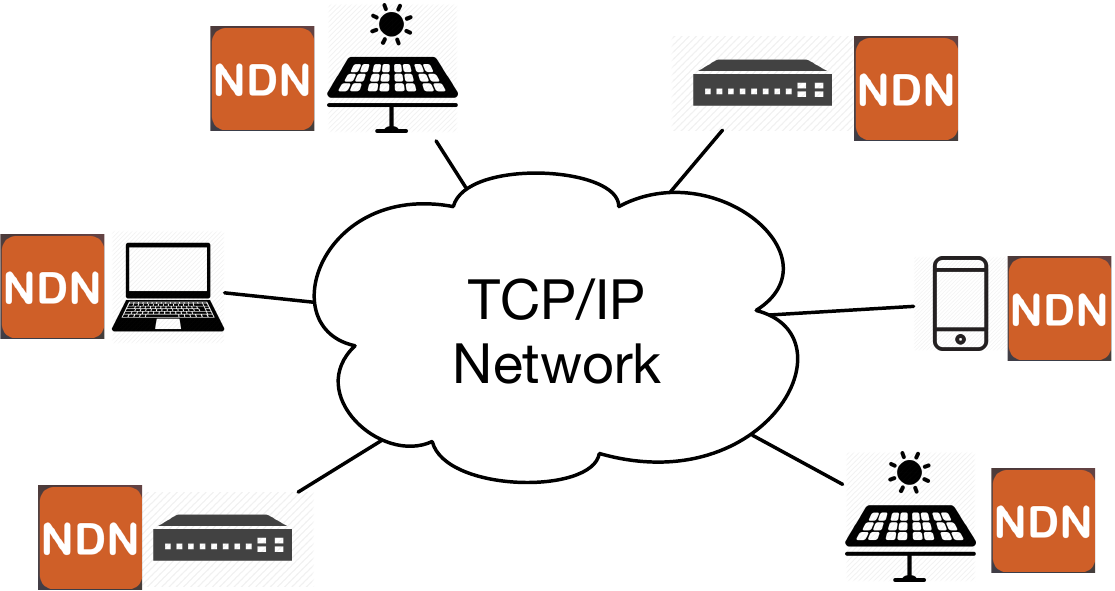}
		\caption{Overlaid NDN on TCP/IP}
		\label{fig:overlay}
	\end{subfigure}
	\caption{Two Different Approaches to DLedger's Network}
\end{figure}

For the sake of better understanding, we use the experimental solar gateway system as an example to explain the network protocols used in DLedger.
The solar gateway system is based on a device-to-device mesh wireless network ( Figure~\ref{fig:mesh}) connecting constrained solar devices with LoRa, which is a typical network scenario in IoT.
Each device in the solar system runs the NDN protocol stack directly over link layer protocols.

%We show that building DLedger over the NDN network in a private system benefits the record dissemination by leveraging NDN's built-in content multicast, thus saving the network bandwidth and lower the latency.
%Moreover, in cases of network failures and other situations when the origin of a new record is offline, NDN helps to get the record from other nodes.

If the business system is in a large scale (wide area network), another approach is to build an overlaying NDN mesh network on top of TCP/IP architecture (Figure~\ref{fig:overlay}), similar to Internet Relay Chat (IRC) over TCP in Bitcoin.
DLedger can function as expected but cannot benefit from NDN in such deployment, where each entity needs to maintain a list of first-hop neighbours' IP addresses and follow the gossip protocol to broadcast like Bitcoin.

\subsection{Named Data Networking}
\label{sec:network:ndn}

NDN directly uses application data name at the network layer for routing and packet forwarding.
Different from an IP packet, neither Interest nor Data packets carry addresses.
To deliver packets in the network, NDN leverages its stateful forwarding plane~\cite{yi2013case}.
When receiving an Interest (e.g., \name{/DLedger/gtw-node0/f46...1ace}), NDN forwarders will forward the packet to corresponding interface(s) by its name and record its path---incoming interfaces and outgoing interfaces.
When the matching Data packet arrives, forwarders will forward it back to the requester following the reverse path taken by the Interest.
At the same time, forwarders will cache the Data packets to satisfy potential Interests in future targeting the same Data.
Getting rid of addresses enables more efficient data dissemination: 
\begin{itemize}  [leftmargin=*, itemsep=1pt, parsep=2pt, topsep=4pt]
	\item \textit{In-network Caching}: Routers cache Data packets to satisfy future Interests,
	\item \textit{Interest Aggregation}: Multiple Interests targeting the same data can be aggregated into one.
\end{itemize}
Thanks to these two features, NDN naturally supports \emph{content multicast}.

To serve data under a specific prefix (e.g., \name{/DLedger}), a node will need to register this name prefix to the network showing its willing to receive such Interest packets.
In DLedger, each entity will register two name prefixes to the network for the wireless interface to receive Interest packets:
\begin{itemize} [leftmargin=*, itemsep=1pt, parsep=2pt, topsep=4pt]
	\item Prefix \name{/DLedger} to receive multicast Interests.
	\item Prefix \name{/DLedger/<entity's name>} to receive unicast Interests, where \name{<entity's name>} is unique per entity.
\end{itemize}

%Since NDN no longer cares where (i.e., IP address) to fetch the target data, the network has more flexibility to fetch the content in a more intelligent way.
%Specifically, NDN uses a smart decision maker called \emph{forwarding strategy} in its forwarding plane to decide how an Interest under a specific name prefix should be forwarded.
%NDN supports \emph{Interest multicast} at the network layer by utilizing multicast strategy and multicast prefix.
%A multicast strategy will forward an Interest packet to all the interfaces where the matched multicast prefixe was registered.
%For example, entities who want to receive multicast Interests under the prefix \name{/DLedger} simply register this prefix to the network, showing their interests in receiving such Interest packets.
%When forwarding an Interest under \name{/DLedger}, a router will send the packet to all the matched interfaces.

To work with data-centric data dissemination, NDN requires data producers to sign each Data packet on creation and encrypt its content if needed.
Therefore, Data packets are secured directly during transmission and storage on untrusted servers.
This method fundamentally differs from today's TLS~\cite{RFC8446} or QUIC~\cite{quic-sigcomm17}, which only protect point-to-point connections.
Using public keys certified by designated identity managers, a node in NDN can ensure data integrity and verify producers' authority through trust chain.
Encryption, on the other hand, ensures confidentiality of Data packets.

Building over NDN's Interest and Data packet exchange, DLedger directly uses an NDN Data packet as a record, thus getting rid of the encapsulation overhead.
As shown in Figure~\ref{fig:ndn-packet} and Figure~\ref{fig:record}, record names are Data names, approved records and payload fields are Data contents, and PoAs are Data signatures.
Especially, a record is structurally named \name{/DLedger/<Generator ID>/<Record hash>}, with the first two conponents revealing where to fetch it.
In the solar network example, the record \name{/Dledger/gtw-device0/f46...1ace} is created by the entity \name{/DLedger/gtw-device0}.
Hence, packet wrapping and network functions are supported by NDN, vastly simplifying DLedger's implementation.

%In general, NDN network is deployable in the private network system either
%\begin{enumerate*} [label=(\roman*)]
%	\item as a pure network stack (i.e., run NDN directly over link layer), which can coexist with TCP/IP stack,
%	or
%	\item as an overlay over the existing TCP/Iit P network.
%\end{enumerate*}
%In fact, to fully leverage NDN's features to facilitate the system data dissemination and security, it is better to use a pure NDN stack.
%As an simple example, NDN's Interest packet reveals no source information, but overlaying TCP/IP potentially leak the requester's IP address.
%We argue that deploying NDN is doable in the private system where the system vendor or the business provider has control over the applications in the whole system.
%For example, in the solar network where solar gateway devices form a meshed ad-hoc network over Long Range wireless (LoRa), each device has a pure NDN stack installed.

\subsection{Notification Protocol}
\label{sec:network:advertisement}

After generating a record, to notify other peers to fetch it , a peer advertises it via a Notification Interest (Notif Interest) with name
\begin{center}
	\name{/DLedger/NOTIF/<Generator ID>/<Record hash>}
\end{center}
Triggered by this Interest, a peer can simply drop the \name{NOTIF} component to compose an Interest for the new record and fetch the new record from the network.

To prevent potential denial-of-service attack and reflection attack by abusing the Notif Interest, the business service provider may require each entity to carry the PoA of the new record when multicasting the Notif Interest.
Since PoA is calculated over the record name, it is sufficient for a Notif Interest receiver to verify this PoA by recovering the record name from the Notif Interest name (by dropping the  \name{NOTIF} component).
Embedding PoA in the Notif Interest will not lead to extra overhead because with PoA already being verified when receiving Notif Interest, there is no need for an entity to double verify the PoA after fetching the actual record Data packet.
Instead, an entity only needs to ensure the record matches the name and the PoA carried in the previous Notif Interest.

\subsection{Synchronization Protocol}
\label{sec:network:sync}
\label{sec:network:merge}

In DLedger, an entity triggers a synchronization process by multicasting a Sync Interest to its neighbors.
The Sync Interest carries the list of tailing records in the node's current view of local ledger as the Interest parameter~\cite{interest-parameter} with the following naming convention:
\begin{center}
	\name{/DLedger/SYNC/<tailing records digest>} \\
\end{center}

%In DLedger, besides listening Notification Interests and fetching new records, an entity can periodically send $I_{sync}$ to double check the freshness of the local ledger
%After a node returns back online from a sleep, is recovered from a network failure, or any other such condition, the Sync Interest should be sent immediately.

When another node receives a Sync Interest $I_{sync}$, it compares the list in $I_{sync}$ with its local list of tailing records, figuring out whether the local ledger or the sender's ledger is out-of-date. 
If there exist tailing records missing locally, for every missing record, it sends an Interest to fetch it.
The ancestor records approved by retrieved records will also be recursively checked and fetched.
On the other hand, if a record in $I_{sync}$ is no longer a tailing record in its local ledger, it will send out an Sync Interest declaring the sender's ledger is out-of-sync.
%
%To fetch a missing record during the synchronization, an entity will multicast the Interest carrying the desired record name to the network.
%The multicast Interest will first try to fetch the missing record from the neighbor peers.
%Since the synchronization is essentially through the tailing record comparison among neighbor peers, at least one neighbor should have the record and reply the Data unless the neighbor peer goes offline.
%It is also possible that more than one neighboring peers will answer the Interest, in the wireless mesh network, proper collision avoidance mechanism should be applied to avoid collisions.
%If none of the currently available neighbors have the desired record, the Interest will be further forwarded to the origin generator of the record.
%%another one hop neighbors, resulting in a hop-by-hop multicast model (Figure~\ref{fig:sync}).
%Eventually, the requester of a record as well as its neighboring nodes, all will receive this missing record and update their local ledgers.

%\subsection{Synchronization after Network Partitions}

Thanks to DAG, synchronization is able to merge multiple ledgers seamlessly after recovering from a network partition.
Instead of leading to forks when using blockchain, aggregations in DLedger exposes no conflicts.
%After reunion of multiple subnets, the periodic synchronization process will recover an entity's view of the whole DAG.
Once synchronized, an entity is allowed to create a record approving the existing records from multiple branches formed by different subnets, thus merging multiple branches together.

Note that the contribution policy won't block the synchronization process because it only applies to tailing record, but the interlock policy will still apply.
If a record fetched during the synchronization made a self-approval, this record will not be approved by good peers in the future.

\subsection{Efficient Data Dissemination in DLedger: A Case Study}

We illustrate how NDN can help DLedger achieve efficient data dissemination in heterogeneous IoT network using the notification process and record fetching process as examples.

\begin{figure} [ht]
	\centering
	\begin{subfigure}{0.3\textwidth}
		\centering
		\includegraphics[width= \textwidth]{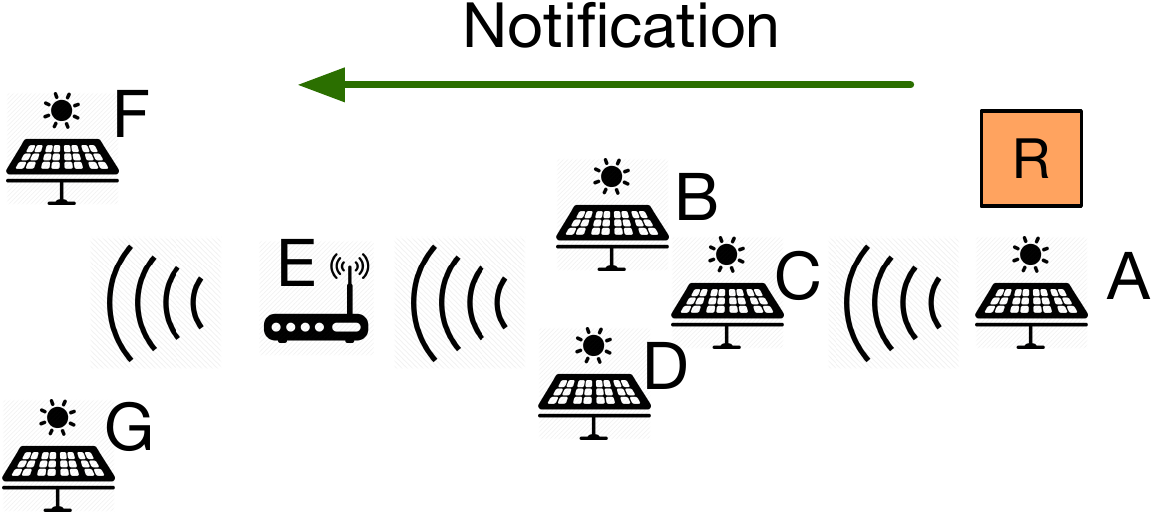}
		\caption{Notification Interest}
		\label{fig:notif}
	\end{subfigure}
	\begin{subfigure}{0.3\textwidth}
		\centering
		\includegraphics[width=\textwidth]{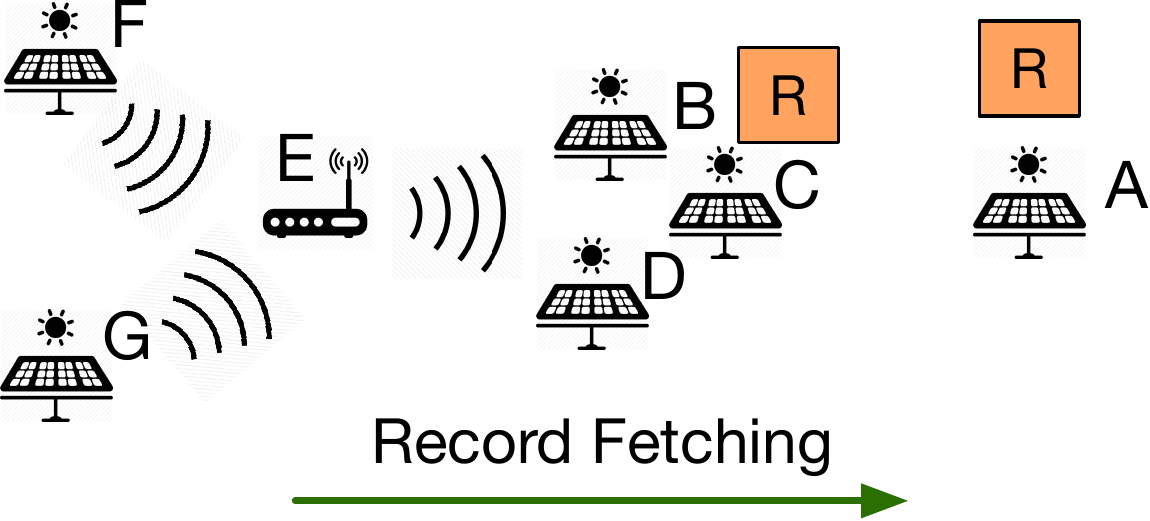}
		\caption{Record Fetching}
		\label{fig:fetch}
	\end{subfigure}
	\caption{Notification and Record Fetching}
\end{figure}

As shown in Figure~\ref{fig:notif}, after the record generator $A$ multicasts the Notif Interest to its first-hop neighbors $B$, $C$, and $D$, they will relay the multicast to the further node $E$. 
Duplicated Interests are suppressed here, i.e., if $B$ or $D$ hear the same Notif Interest multicasted by $C$, they can avoid sending the packet again.
This feature is absent in the muticast solutions based on IP or other network protocols (e.g., Zigbee, Thread, etc.), because $B$, $C$, or $D$ cannot recognize the content carried in the multicast packet at the network layer.

When entities hear a notification, they will send Interests for the same Data in a short time, thus those Interests can be aggregated by the intermediate nodes and the replied Data can be cached to satisfy future Interests, saving the network bandwidth.
For example, as shown in Figure~\ref{fig:fetch}, assuming $C$ has the new record, when $F$ and $G$ wants to fetch it, their two Interests can be merged at $E$ and satisfied by the content store of $C$ instead of $A$.

Interests can be retransmitted in case of intermittent connectivity or packet loss.
For example, when $C$ is returning the record to $F$ and $G$, the link between $E$ and $F$, $G$ becomes unavailable.
Then, when the connectivity recovers, $F$ and $G$ can retransmit the fetching Interest and get Data from the cache on $E$ even though $E$ is only a forwarder without running the ledger. 

%% file: 5-sec-assess.tex
\section{Security Assessment of DLedger System}
\label{sec:assessment}

This section presents the security assessment of DLedger in terms of its security attributes and how our design can mitigate various attack scenarios and potential vulnerabilities.

\subsection{Security Attributes}

\begin{table}[h]
	\centering
	\caption{Security Attributes of DLedger}
	\begin{tabular}{|| c | m{4.6cm}||} 
		\hline
		\textbf{Security Attribute} & \textbf{DLedger's Approach} \\
		\hline\hline 
		Availability & Decentralized replication of the ledger\\
		\hline
		Integrity & PoA makes confirmed records immutable \\ 
		\hline
		Authenticity & PoA with all the issued/revoked certificates appended in the distributed ledger \\ 
		\hline 
		Confidentiality & Encrypted Content which is invisible to non-peers entities\\ 
		\hline 
	\end{tabular}
	\label{table:security}
\end{table}

Table~\ref{table:security} summarizes the security attributes provided and briefly explains DLedger's methodology.
DLedger provides the first three security attributes by the distributed ledger and PoA.

Hiding entities' identities using pseudonymity is not the main consideration in the design of DLedger.
First, DLedger is designed for private business network where the identity manager, or the business service provider, has the knowledge of their customers' identities.
Second, different from the pseudonymity in Bitcoin where miners are anonymous (i.e., hide behind the arbitrary wallet addresses), an entity and all its records are associated with its PoA.
Moreover, as pointed by many works~\cite{reid2013analysis, miers2013zerocoin}, in Bitcoin and other cryptocurrency systems, pseudonomity can be easily compromised when attackers can combine off-network information with the openness of transactions.

However, since the use case is private business systems, internal records should not be exposed to the public Internet.
To protect data from leakage to the external network, entities can encrypt the content.
This requires the deployment of proper key management and access control.
We argue that this is relatively easy in a private business model where the trust relationships among system entities have already been established.
For example, DLedger can utilize name-based access control (NAC)~\cite{zhang2018nac}, where the identity manager can serve as the decryption key distributor who grants the access rights to internal entities only.

\subsection{Threats Mitigation}

\subsubsection{Laziness}

DLedger utilizes its security policies to demotivate peers from being lazy.
For example, if a peer approves records that have already been confirmed, other peers will abandon them by the contribution policy.
Since certificate revocations are also appended to ledger, it is important for all peers to synchronize, get up-to-date security information, and avoid approving a revoked peer's records. 

\subsubsection{Spam Record Attack}
\label{sec:assess:spam}
PoA is efficient even for constrained devices, enabling a malicious node to fabricate spam records with a valid PoA.
Such spam attack could abuse the system by indefinitely increasing the unconfirmed records (i.e., the hops from a tailing record to a confirmed record) and exploiting legit peers' resources for verification and storage.
It will also increase the network overhead and induce latency into the network.
DLedger prevents such abuse from its design because
\begin{enumerate*} [label=(\roman*)]
	\item With PoA revealing the creator's identity, the attacker will leave its footprints when attacking.
	\item The interlock policy prevents an attacker adding the unconfirmed depth by approving itself.
\end{enumerate*} 
Moreover, application level semantics can help mitigate the abuse.
For example, a node could measure the rate of incoming Notification Interests from each peer to detect malicious nodes, and then report them to the service provider for further examination.

\subsubsection{Denial-of-Service Attack and Reflection Attack}

Since a Notif Interest will trigger system entities to send packets towards \name{/DLedger/Generator ID/<Record hash>} without protection, misbehaving entities or external attackers may abuse this Interest for denial-of-service attack and reflection attack by forging nonexistent \name{<Generator ID>} and \name{<Record hash>} or using a victim's \name{<Generator ID>}.

As mentioned in Section~\ref{sec:network:advertisement}, the system can force each entity to append its PoA into Notif Interests.
In this way, if the PoA is unverifiable or mismatches with the generator's name, the receiver should reject the Interest.
A peer can also report such Interests to an intrusion detection system (IDS) if the business provider deployed one.

\subsubsection{Collusion Attack}

The collusion attack is possible when multiple entities in the P2P system want to maximize their benefits through approving and validating each other's invalid records.
PoA can only ensure data authenticity but not prevent a malicious entity from abusing its approvals.
And neither the interlock policy nor the contribution policy can stop colluded peers collaborating.

DLedger prevents the collusion attack to the extent that unless $W_{confirm}$ or more entities collude, the attack cannot succeed.
Particularly, a record needs to have an weight of at least $W_{confirm}$ to be confirmed.
Thus, faked records cannot get the system's consensus on their validities as far as the number of colluded peers do not exceed $W_{confirm}$.

\subsubsection{Identity Manager's Bot Attack}

All the peers in DLedger trust designated identity managers and certificates they issue.
These designated identity managers might show a malicious behavior by spinning many virtual nodes and assigning them valid identity certificates.
In this way, these virtual nodes can easily confirm invalid records by making $W_{confirm}$ virtual nodes to approve them.

However, DLedger's security by publicity prevents an identity manager from abusing its authority.
As noted in section~\ref{sec:poa}, identity managers must issue certificates by appending them into the ledger system.
This mechanism required all interventions from identity managers to be recorded, allowing later examination if suspicious approvals are identified.

\subsubsection{Light Node Vulnerability}
%
%In traditional distributed ledger systems over TCP/IP, since most of the peer devices don't have enough storage capacity to store entire ledger and perform ledger walks for transaction verification, nodes are differentiated based on their resource and storage capacity into full nodes and light nodes.
In distributed ledger systems, light nodes themselves don't verify blocks or store a copy of the ledger, but rather outsource it to full nodes, causing severe security vulnerabilities.
Taking BitCoin's light node~\cite{btc-light} as an example:
light nodes rely on full node to validate the transactions; consequently, without the help of full node, it accepts fake or invalid coins, resulting in this node's bankruptcy.
Even worse, a light node has no privacy because it sends the wallet address to and receives the balance and transaction history from the full node it depends.

DLedger mitigates this vulnerability by making it easier for every entity to ``mine" new records and preserve a local ledger.
Here, PoA is lightweight even for constraint IoT devices, thus encouraging entities to perform their own computations rather than leasing out.
Also, DLedger's decentralized archiving mechanism as described in Section~\ref{sec:design:archive} allows efficient use of storage capacity without loss of any required ledger data.
NDN, as the underlying network, also helps to address the challenge because any record can be retrieved from any node rather than some specific full nodes, without revealing the identity of the requester.

%% file: 6-imple.tex
\section{Evaluation of DLedger}
\label{sec:impl}

In this section, we evaluate DLedger's robustness, scalability, and ability of handling network partition through the theory analysis and our simulation results over ndnSIM~\cite{mastorakis2017ndnsim}, an NS-3 based simulation platform for NDN.

\subsection{The Unconfirmed Records Size}

An essential concern of DLedger is its robustness: as system runs, will the size of unconfirmed records go infinitely large resulting in the system crash?
As proved in Appendix~\ref{ap1}, the tailing record size is concentrated around the constant:
\begin{align*}
\label{eq:tailing-size}
	\frac{n \lambda T}{n-1}
\end{align*}
where $T$ is the average latency for a new record to be accepted by the system and $\lambda$ is the new record generation rate of the system following a Poisson distribution.
Moreover, Appendix~\ref{ap2} shows that the time for a record being confirmed is also a constant value.
Therefore, since both the frontier (tailing record) and the depth (time to be confirmed) of DAG's unconfirmed records keep constant, the unconfirmed record size is also a constant value.

\begin{figure} [ht]
		\centering
		\includegraphics[width= 0.35\textwidth]{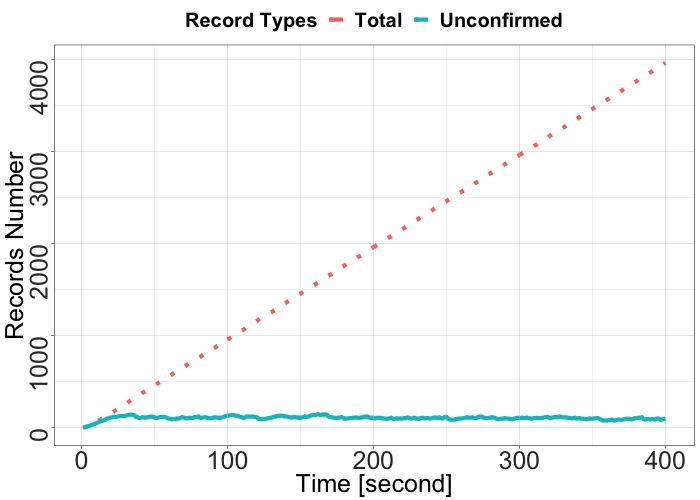}
		{\footnotesize \\
			\begin{flushleft}
			Simulation Settings: 50 entities P2P network with $W_{confirm} = 20$. Each entity generates a new record every 5 seconds. \par
			\end{flushleft}}
		\vspace{-3mm}
		\caption{Unconfirmed Records As DAG Grows}
		\label{fig:unconfirmed-size}
\end{figure}

We evaluate DLedger by the simulation in terms of the unconfirmed record size as the DAG grows.
To be specific, we set up a 50 entities P2P network with $W_{confirm} = 20$, where each entity's record generation rate is 0.2 records/sec.
As shown in the Figure~\ref{fig:unconfirmed-size}, the unconfirmed record number will not increase with the DAG's growth.
Instead, the unconfirmed record number line will increase for a short time after bootstrapping, and then fluctuate around a constant value, confirming the mathematics proof.

\begin{figure} [ht]
	\centering
	\includegraphics[width=0.35\textwidth]{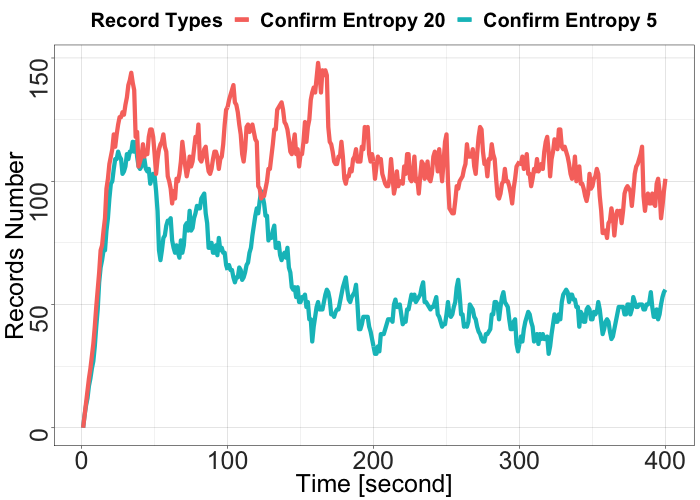}
	{\footnotesize \\
		\begin{flushleft}
		Simulation Settings: 50 entities P2P network with $W_{confirm} = 5$ and $W_{confirm} = 20$. Each entity generates a new record every 5 seconds. \par
		\end{flushleft}}
	\vspace{-3mm}
	\caption{The Unconfirmed Records With Different Weight}
	\label{fig:size-weight}
\end{figure}

The Figure~\ref{fig:size-weight} reveals the relationship between the $W_{confirm}$ setting and the unconfirmed record size when the total entity number is the same: 
by adjusting the $W_{confirm}$ from 5 to 20, the unconfirmed record size will increase by only about 100\%.

\subsection{System Scalability}

We argue that DLedger has better scalability because of the following two reasons.
First, DLedger leverages DAG to keep the records.
Instead of allowing only one successful next block in the blockchain, DLedger accept multiple valid next records, thus allowing multiple entities contributing to the system in parallel.
This gets rid of the computation waste and frees the limit on record processing speed.
Second, PoA is used as the gating function.
Unlike PoW which is highly CPU bound, PoA is much cheaper and efficient even for constrained IoT devices, greatly improving the system efficiency.

\TODO{Zhiyi: do this simulation for 5 times and make the plot more scientific.}

\begin{figure} [ht]
	\centering
	\includegraphics[width= 0.7\linewidth]{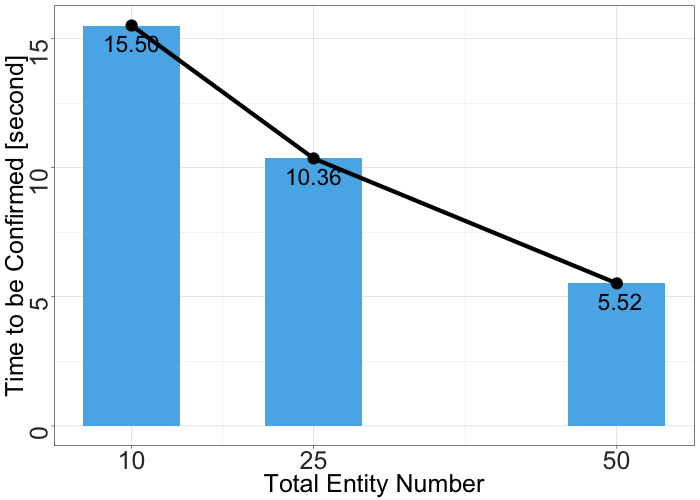}
	{\footnotesize \\
		\begin{flushleft}
		Simulation Settings: $W_{confirm} = 5$ with total entity number 10, 25, and 50.\par
	    \end{flushleft}
		}
	\vspace{-3mm}
	\caption{Unconfirmed Records As DAG Grows}
	\label{fig:scalability}
\end{figure}

We also show DLedger's performance when the number of entities grows.
As shown in Figure~\ref{fig:scalability}, with fixed $W_{confirm}$, the more entities participate, the faster a record will be confirmed.
Because for any specific record, more entities means more efforts on approving it.

\subsection{Network Partition}

\begin{figure} [t]
	\centering
	\includegraphics[width= 1\linewidth]{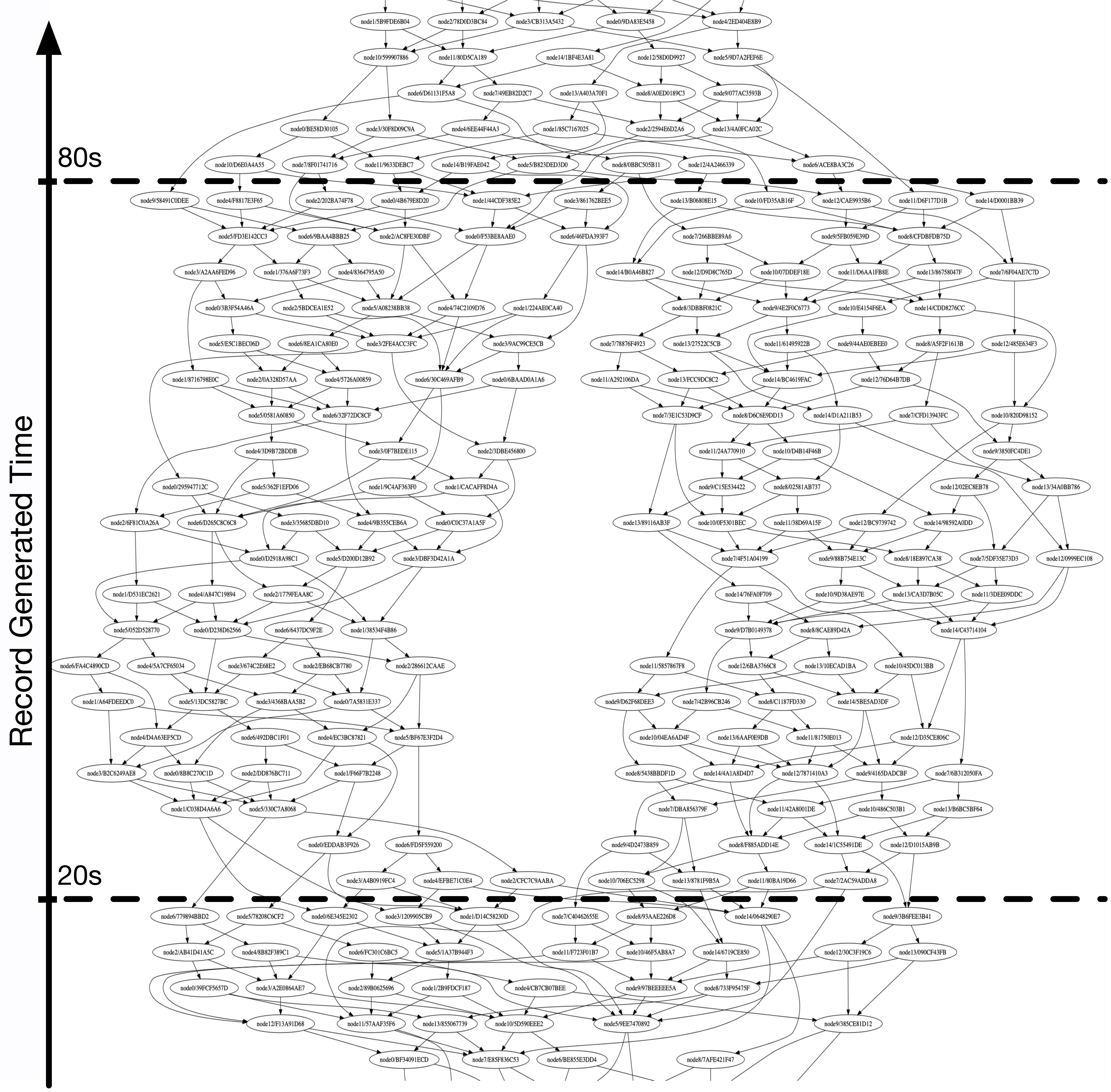}
	{\footnotesize \\
		\begin{flushleft}
		This shows a DAG from the simulation. Each circle in the figure is a record. The links among records are the approvals.
		The recent records are higher in the DAG as shown in the figure. \\
		Simulation Settings: 15 entities P2P network. 
		The partition takes place at second 20, dividing the network into a 7-node subnet and a 8-node subnet. Two networks rejoin at 80 seconds. \par
	\end{flushleft}}
	\vspace{-3mm}
	\caption{Unconfirmed Records As DAG Grows}
	\label{fig:network-partition}
\end{figure}

We evaluate DLedger's performance in the case of network partition: in the simulation, we split the network into two independent subnets by taking down the links between them.
The partition takes 100 seconds and two networks unit again after that.
Figure~\ref{fig:network-partition} shows a DAG maintained by an entity at the end of the simulation.
As shown, two branches have formed and then merged, showing DLedger's ability to recover from the network partition, eventually confirming records from each subnet.

%% file: 7-discussion.tex
\begin{table*}[t]
	\centering
	\caption{Comparison Between PoA and Other Consensus Algorithms}
	\begin{tabular}{||c c c c c ||} 
		\hline
		Gating Function & Consensus & IoT Friendly & Identity Management & Applied Scenarios \\ [0.5ex] 
		\hline\hline
		Proof-of-Work & Yes & No  & Not needed & Public System \\ 
		\hline 
		Proof-of-Stack  & Yes & No & Not needed & Public System \\ 
		\hline 
		Proof-of-Space & Yes & No & Not needed & Public System \\ 
		\hline 
		Proof-of-Activity & Yes & No  & Not needed & Public System \\
		\hline 
		Proof-of-Assignment* & Yes & Yes  & Needed & Public System \\
		\hline
		Proof-of-Authentication (PoA) & Work with Weight  & Yes & Needed & Private System \\ 
		\hline
	\end{tabular}
	\vspace{2mm}
	{\footnotesize \\
		*: As discussed in Section~\ref{sec:observation},  Proof-of-Assignment subjects to single-point-of-failure and attack scenario where attackers can hire mild computation power.\par}
	\label{table:poa-pow}
\end{table*}

\section{Discussion}
\label{sec:discussion}

\subsection{System Bootstrapping}

The business service provider should bootstrap DLedger for entities to use.
The service provider needs to generate the first several records as the genesis records for later records to approve.
System constants, such as $E_{confirm}$ and $E_{contribute}$, should also be properly configured.
These configurations can be dynamically tuned via software updates, which can be delivered through records in DLedger.
After the bootstrapping, the identity manager can go offline unless when new entities join or certificate revocation operations are needed.

To join the DLedger P2P network, a new entity should have a certificate issued by the identity manager.
The issuance is done by inserting a record into DLedger, making other peers aware.
Then the new entity can start working with a synchronization.

\subsection{PoA v.s. Other Consensus Algorithm}

Compared with the consensus algorithms used and proposed by other distributed ledger systems, PoA is not like a consensus algorithm.
Actually, PoA only associates an entity's public identity with the records it generates and ensures the data integrity.
This explains why PoA is much cheaper than any other consensus algorithms.
In DLedger, the consensus on a confirmed record is achieved by relying on enough number of different entities to approve it, where PoA plays a key role to distinguish how many different entities have approved that record.

This is also the fundamental difference between IOTA and DLedger.
In IOTA, the miner is anonymous and thus enough approvals do not necessarily mean enough number of entities in the system recognize the validity of a record.

Another notable difference between PoA and other consensus algorithms is that PoA requires the identity managers to certify the peers in the system.
This prerequisite decides that DLedger is suitable for private business systems such as bank accounting, medical records and power grid systems, but not for public network systems.
The comparison between PoA and other consensus algorithms is shown in the Table~\ref{table:poa-pow}.

\subsection{Record Snapshot}
Section~\ref{sec:design:archive} presents the mechanism for an entity to reduce the size of its local ledger.
In this section, we discuss another mechanism called record snapshot when the data from multiple records is aggregatable and thus can be summarized into one record.
For example, in the solar gateway network, the records keep each entity's residential power production/consumption to get the power balance.
In this case, a snapshot process can be performed by calculating the eventual balance for each entity using the confirmed records so that these confirmed records can be dropped to save space.
Note that taking snapshot is an individual behavior without needing to synchronize snapshot records with other peers.

\subsection{Ledger Records For Cert Revocation}
Besides application records inserted, as mentioned, DLedger also keeps the issuance and revocation of certificates.
This section illustrates how to revoke a certificate.

When an intrusion such as spam attack is detected, the identity manager may revoke the certificate of the malicious node by signing a notice of revocation and attaching it to DLedger.
Similar to generating regular records, the identity manager needs to verify $n$ existing records and multicast a notification to the whole P2P network.

\begin{figure} [ht]
	\centering
	\includegraphics[width= 0.6\linewidth]{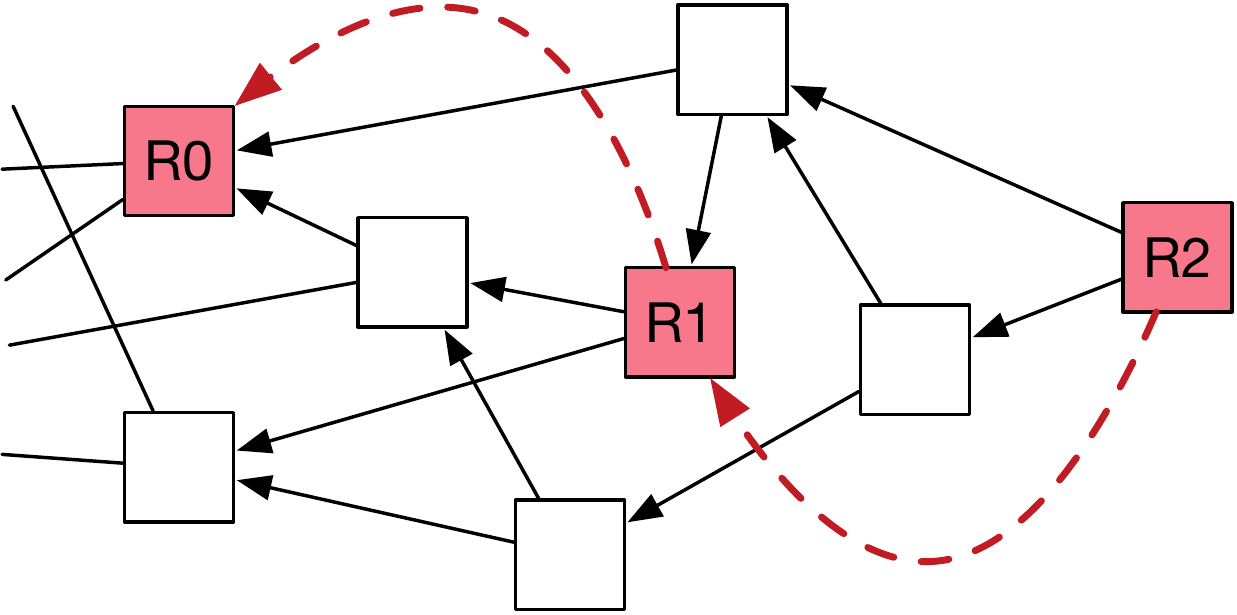}
	\caption{Attaching Notice of Revocation}
	\label{fig:revocation}
\end{figure}

To facilitate querying the revocation status, the overlay pointers can be applied.
For example, as shown in figure~\ref{fig:revocation}, $R1$ and $R2$ are DLedger records carrying notices of revocation.
Besides approvals, they also have an application-level pointer to the previous revocation in the payload, accelerating search for revocation.
In the example, nodes only need to maintain the last revocation record $R2$ and use it to follow back to all revocations.

\subsection{Other Use Cases of DLedger}
The DLedger design shown in the paper serves two different use cases:
\begin{enumerate*} [label=(\roman*)]
	\item application record ledger
	and
	\item identity manager's certificate management.
\end{enumerate*}

The DLedger can also be used to facilitate IoT systems in terms of configurations and information sharing.
As a simple example, when building DLedger over TCP/IP networking, the system needs to maintain the mapping between NDN name prefixes and IP addresses.
Similar to how Bitcoin's IRC overlay maintains neighbors' IP addresses, DLedger can be utilized here.
For example, when an entity joins the P2P network, it can insert a record containing its address.
Whenever its IP address changes, it appends a record containing a pointer to obsolete the stale record.

%% file: 9-conclusion.tex
\section{Conclusion}
\label{sec:conclusion}

We present DLedger, a distributed ledger system, to satisfy the need for IoT-friendly ledger in private business scenarios.
Utilizing DAG and PoA, even constrained devices can engage by ``mining" their own records and getting them confirmed, which enhances data availability, integrity and authenticity.
Building over NDN further enables effective data dissemination in unstable IoT networks and simplifies implementation.

Through designing DLedger, we want to prove the power of distributed ledgers differently from existing solutions.
That is, instead of combining data openness with anonymity or pseudonomity, DLedger exploits verifiable identities known within the private system, offering an explicit way to measure the validity of a record by the number of approvers.
Associating records with their generators also push entities into behaving good by recording the footprints of attackers.
Hence, even with great authority, the service provider cannot freely add bots because the certificates of them will be public to the whole system.
It is such a reconsideration on the methodology that simplifies the achievement of security, without hurting IoT-friendliness and efficiency.

%% file: A1-math.tex
\appendix
\section{DLedger's Robustness Math Proof}

\subsection{Constant Tailing Record Size}
\label{ap1}
We assume that at any time $t$, the tailing record size $g(t)$ in DAG remains stable and we assume $g(t)$ is concentrated around the constant $C$.
In DLedger, we let the entities' record appending to be a Poisson process with a rate $\lambda$.
We also consider the network latency in the system: after a record being created at time $t$, it takes $T$ time to be visible to the entire system (at time $t + T$).

In our proof, we let all the entities in the system behave as expected, i.e., each entity randomly select $n$ existing tailing records out of total from $C$ tailing records from their local DAG to approve when creating a new record.
However, at any time, since there are $\lambda T$ tailing records that are still not visible, by the assumption, there should also be approximately $\lambda T$ tailing records becoming approved.
Therefore, from the system perspective, the number of nodes which remain tailing records is actually $C - \lambda h$.
Hence, the probability of an approved record being tailing record is:
\begin{gather*}
	\frac{C - \lambda T}{C}
\end{gather*}

By the stationarity assumption of the tailing record size, we have 
\begin{gather*} 
	\lambda T = n \lambda T \frac{C - \lambda T}{C}
\end{gather*}
Solving the equation, we have:
\begin{gather*}
C = \frac{n \lambda T}{n-1}
\end{gather*}

\subsection{The Confirmation Time}
\label{ap2}
Assume there are totally $N$ entities in the system.
Let $f(W_{confirm}, y)$ denote the expected approvals for a node whose weight is $y$ to be confirmed ($w=W_{confirm}$).
That is, on average, a record whose $w=y$ still needs to be further approved by $f(W_{confirm}, y)$ approvals to satisfy the weight condition and get confirmed.
Obviously, we have $f(W_{confirm}, W_{confirm})=0$ and $f(W_{confirm}, 0)$ is the total required number of approvals for a record to be confirmed.

For a record with $w=y$, the probability of a new approval contributing to the weight is $\frac{N-y}{N}$ and $\frac{y}{N}$ on the contrary, which results in the equation:
\begin{gather*}
	 f(W_{confirm}, y)= \\
	 f(W_{confirm}, y)\frac{y}{N} + f(W_{confirm}, y+1)\frac{N-y}{N} + 1
\end{gather*}
By solving the equation, we can get 
\begin{gather*}
f(W_{confirm}, 0)=N\sum_{i=0}^{W_{confirm}-1}\frac{1}{N-i}
\end{gather*}
which is the expectation of approval number for a record to be confirmed.

Since the tailing record number is constant around $C$ and for each node, it takes a constant number of approvals to get confirmed.
Because an entity randomly selects the existing tailing records to approve when generating a new record, the time for a record to get $f(W_{confirm}, 0)$ number of approvals should also be a constant value.

We can get the upper bound of the confirmation time by the fact that for any unconfirmed node there are two cases: either it is a tailing record, or there exists at least one tailing record which approves it directly or indirectly. 
Given the average time for a tailing record to be approved is:
\begin{align*} 
t_{approved} = T+\frac{C}{n\lambda} \\
=\frac{n}{n-1}T
\end{align*}
the expected time for a new node to be confirmed is less than 
\begin{align*} 
t_{confirm} \leq f(W_{confirm}, 0) \times t_{approval} \\
 = \frac{TNn}{n-1}\sum_{i=0}^{W_{confirm}-1}\frac{1}{N-i}
\end{align*}